\documentclass{amsart}

\usepackage[english]{babel}
\usepackage[utf8]{inputenc}
\usepackage[T1]{fontenc}

\usepackage{amsmath}
\usepackage{amssymb}
\usepackage{mathtools}
\usepackage{amsthm}
\usepackage{physics}
\usepackage{cite}
\usepackage{enumerate}

\usepackage{booktabs}
\usepackage{rotating}
\usepackage{tabularx}
\usepackage{multirow}

\usepackage{siunitx}

\usepackage{graphicx}
\usepackage{subfig}
\usepackage{wrapfig}
\usepackage{xcolor}

\usepackage{etoolbox} 
\AtBeginEnvironment{pmatrix}{\everymath{\displaystyle}}
\AtBeginEnvironment{bmatrix}{\everymath{\displaystyle}}

\DeclareMathOperator{\dpat}{dp}

\newcommand{\Ilow}{I_\text{low}}
\newcommand{\Ihigh}{I_\text{high}}
\newcommand{\Pj}{\mathbb{CP}}
\newcommand{\Cp}{\mathbb{C}}
\newcommand{\dP}{\mathrm{d}P}
\newcommand{\pvec}[1]{\mathbf{[#1]}}
\renewcommand{\vec}[1]{\mathbf{#1}}
\DeclareMathOperator{\ud}{d}

\newcounter{eqlist}

\newcounter{eqlistred}

\allowdisplaybreaks
\makeatletter 
\renewcommand{\pdv}[2]{\begingroup 
\@tempswafalse\toks@={}\count@=\z@ 
\@for\next:=#2\do 
{\expandafter\check@var\next\@nil
 \advance\count@\der@exp 
 \if@tempswa 
   \toks@=\expandafter{\the\toks@\,}%
 \else 
   \@tempswatrue 
 \fi 
 \toks@=\expandafter{\the\expandafter\toks@\expandafter\partial\der@var}}%
\frac{\partial\ifnum\count@=\@ne\else^{\number\count@}\fi#1}{\the\toks@}%
\endgroup} 
\def\check@var{\@ifstar{\mult@var}{\one@var}} 
\def\mult@var#1#2\@nil{\def\der@var{#2^{#1}}\def\der@exp{#1}} 
\def\one@var#1\@nil{\def\der@var{#1}\chardef\der@exp\@ne} 
\makeatother

\theoremstyle{plain}

\theoremstyle{definition}
\newtheorem{definition}{Definition}
\theoremstyle{remark}
\newtheorem{remark}{Remark}
\theoremstyle{remark}
\newtheorem{example}{Example}

\begin{document}
\title[Complexity \& integrability in 4D bi-rational maps with 2 invariants]{%
    Complexity and integrability in 4D bi-rational maps with two invariants}
\date{\today}
\author[G. Gubbiotti]{Giorgio Gubbiotti}
\author[N. Joshi]{Nalini Joshi}
\author[D. T. Tran]{Dinh Thi Tran}
\address{School  of  Mathematics  and  Statistics  F07,  The  University  of  Sydney,  NSW  2006, Australia}
\email{giorgio.gubbiotti@sydney.edu.au}
\email{nalini.joshi@sydney.edu.au}
\email{dinhthi.tran@sydney.edu.au}
\author[C-M. Viallet]{Claude-Michel Viallet}
\address{LPTHE, UMR 7589 Centre National de la Recherche Scientifique \& UPMC Sorbonne Universit\'e, 4 place Jussieu, 75252 Paris Cedex 05, France}
\email{claude.viallet@upmc.fr}

\begin{abstract}
    In this letter we give fourth-order autonomous recurrence
    relations with two invariants, whose degree growth is cubic or exponential.
    These examples contradict the common belief  that  maps with sufficiently 
    many invariants can have at most quadratic growth.
    Cubic growth may reflect the existence of 
    non-elliptic fibrations of invariants, 
    whereas we conjecture that the exponentially growing cases
    lack the necessary conditions for the applicability
    of the discrete Liouville theorem.
\end{abstract}

\maketitle

\section{Introduction}

Bi-rational maps in two dimensions have played a crucial role in the study
of integrable discrete dynamical systems since the seminal paper of 
\cite{PenroseSmith1981} and the introduction of the QRT mappings in 
\cite{QRT1988,QRT1989}.
Elliptic curves and rational elliptic surfaces proved to be one of the  main
tools in understanding the geometry behind this kind of integrability, 
see \cite{Sakai2001,Duistermaat2011book,Tsuda2004}.
In this letter we give examples of higher-order maps whose properties 
go beyond those of the two-dimensional maps, and  show that the geometry of elliptic
fibrations is no longer sufficient to explain their behaviour.

Up to now the QRT mappings appear to describe almost 
the totality of the known integrable examples in dimension two with 
some notable exceptions \cite{VialletRamaniGrammaticos2004,Duistermaat2011book}. However,  no general framework exists for higher order  maps.
A generalization of the QRT scheme \cite{QRT1988,QRT1989}
in dimension four was given in
\cite{CapelSahadevan2001}. 
Certain maps obtained in \cite{CapelSahadevan2001}
were shown in \cite{Hay2007} to be autonomous reductions of members of
$q$-Painlev\'e hierarchies (\emph{multiplicative equations} in Sakai's scheme \cite{Sakai2001}).
Since hierarchies are known also for the  \emph{additive}
discrete Painlev\'e
equations \cite{CresswellJoshi1999},  it is clear that the cases considered in 
\cite{CapelSahadevan2001} cannot exhaust all the possible integrable 
autonomous maps in four dimensions, as already shown in \cite{JoshiViallet2017}.
It is important to mention that there are  also other examples of discrete mappings of higher orders produced either by periodic or symmetry reductions of 
integrable partial difference equations
\cite{PapageorgiouNijhoff1990,QuispelCapel1991,vanderKampQuispel2010,LeviWinternitz2006R}
or as Kahan-Hirota-Kimura discretization \cite{Kahan1993,HirotaKimura2000} of 
continuous integrable systems 
\cite{PetreraSuris2010,CelledoniMcLachlanOwrenQuispel2013,
CelledoniMcLachlanOwrenQuispel2014,PetreraPfadlerSuris2009}.

In this letter, we focus on the study of integrability
properties of autonomous recurrence relations.
Here an autonomous recurrence relation is given
by bi-rational map of the complex projective space into itself:
\begin{equation}
    \varphi \colon \pvec{x}\in\Pj^{n} \to \pvec{x'} \in \Pj^{n},
    \label{eq:mapp4}
\end{equation}
where $n>1$\footnote{Bi-rational maps in $\Pj^{1}$ are just M\"obius
transformations so everything is trivial.}.
We take $\pvec{x}=\left[ x_{1}:x_{2}:\dots:x_{n+1} \right]$ 
and $\pvec{x'}=\left[ x_{1}':x_{2}':\dots:x_{n+1}' \right]$
to be homogeneous coordinates on $\Pj^{n}$.
Moreover we recall that a bi-rational map is a rational map
$\varphi\colon V\to W$ of algebraic varieties $V$ and $W$ such that there exists a
map $\psi\colon W\to V$, which is the \emph{inverse} of $\varphi$ in the
dense subset where both maps are defined \cite{Shafarevich1994}.

Integrability for autonomous recurrence relations (discrete equations) 
can be characterized in different ways.
In the continuous case, for finite dimensional systems, integrability is
usually understood as the existence of a ``sufficiently'' high number of 
\emph{first integrals}, i.e. of \emph{non-trivial} functions constant
along the solution of the differential system.
In the Hamiltonian setting a characterization of integrability was given
by Liouville \cite{Liouville1855}.
In the case of maps \eqref{eq:mapp4} the analogue of 
first integrals are the \emph{invariants}. 
To be more precise we state the following:
\begin{definition}
    An invariant of a bi-rational map $\varphi\colon\Pj^{n}\to\Pj^{n}$
    is a homogeneous function $I \colon \Pj^{n}\to \Cp$ such that
    it is left unaltered by action of the map, i.e. 
    \begin{equation}
        \varphi^{*}\left( I \right) = I,
        \label{eq:fintdef}
    \end{equation}
    where $\varphi^{*}\left( I \right)$ means the pullback of $I$
    through the map $\varphi$, i.e. 
    $\varphi^{*}\left( I \right)=I\left( \varphi\left( \pvec{x} \right) \right)$.
    For $n>1$, an invariant is said to be \emph{non-degenerate} if:
    \begin{equation}
        \pdv{I}{x_{1}}\pdv{I}{x_{n}}\neq 0.
        \label{eq:nondegcond}
    \end{equation}
    Otherwise an invariant is said to be \emph{degenerate}.
    \label{def:inv}
\end{definition}


In what follows we will concentrate on a particular class 
of invariants:

\begin{definition}
    An invariant $I$ is said to be \emph{polynomial}, if in the
    affine chart $\left[ x_{1}:\dots:x_{n}:1 \right]$ the function
    $I$ is a polynomial function.
    \label{def:polyinv}
\end{definition}

A polynomial invariant in the sense of definition \ref{def:polyinv}
written in homogeneous variables is always a rational function
homogeneous of degree 0.
The form of the polynomial invariant in homogeneous coordinates
is then given by:
\begin{equation}
    I\left( \pvec{x} \right) =\frac{ I'\left( \pvec{x} \right)}{t^{d}},
    \quad d = \deg I'\left( \pvec{x}\right),
    \label{eq:polyinvrat}
\end{equation}
where $\deg$ is the total degree.

To better characterize the properties of these invariants we introduce
the following:
\begin{definition}
    Given a polynomial function $F\colon \Pj^{n}\to V$, where $V$
    can be either $\Pj^{n}$ or $\Cp$, we define the \emph{degree pattern} 
    of $F$ to be:
    \begin{equation}
        \dpat F = \left( \deg_{x_{1}} F,\deg_{x_{2}} F,\dots,\deg_{x_{n}} F\right).
        \label{eq:dpatt}
    \end{equation}
    \label{def:dpat}
\end{definition}

\begin{remark}
    The degree pattern of a polynomial function $F$ is not invariant
    under general bi-rational transformations.
    However, the degree pattern of a polynomial function $F$ is invariant
    under scaling and translations, 
    which are transformations of the form:
    \begin{equation}
        \chi \colon \pvec{x} \to \left[ a \vec{x}+b \right],
        \quad
        a\in\Cp\setminus\{0\},b\in\Pj^{n}.
        \label{eq:scaltrasl}
    \end{equation}
    \label{rem:dpat}
\end{remark}

\begin{example}
    Consider the following map in $\Pj^{2}$:
    \begin{equation}
        \varphi\colon [ x:y:t] \mapsto [-y ( {x}^{2}-{t}^{2})+2 ax{t}^{2}: x ( {x}^{2}-{t}^{2})   : t ( {x}^{2}-{t}^{2})].
        \label{eq:mcm}
    \end{equation}
    This map is known as the \emph{McMillan map} \cite{McMillan1971} and
    possesses the following invariant:
    \begin{equation}
        t^{4} I_{\text{McM}} = x^2y^2+( x^2+y^2-2 a x y ) t^2.
        \label{eq:intmcm}
    \end{equation}
    We have $\dpat I_{\text{McM}}=\left( 2,2 \right)$, i.e.
    it is a \emph{bi-quadratic} polynomial.
    We also note that the invariant of a QRT map \cite{QRT1988,QRT1989}, $I_\text{QRT}$, 
    which is a generalization of the McMillan map \eqref{eq:mcm},
    is the ratio of two \emph{bi-quadratics} in the dynamical variables of $\Pj^{2}$.
    Hence QRT mappings leave invariant a pencil of curves
    of degree pattern $\left( 2,2 \right)$.
\end{example}

\begin{example}
    The invariants of the maps presented in
    \cite{CapelSahadevan2001}, $I_\text{CS}$, are 
    are ratios of \emph{bi-quadratics} in all the four dynamical variables
    of $\Pj^{4}$, i.e. ratios of polynomial of degree pattern $\left( 2,2,2,2 \right)$. 
    In this sense the classification of \cite{CapelSahadevan2001}
    is an extension of the one in \cite{QRT1988,QRT1989}.
\end{example}

Finally we will consider invariants are not of the most general kind,
but satisfy the  following condition.
\begin{definition}
    We say that a invariant $I\colon\Pj^{n}\to\Cp$
    is \emph{symmetric} if it is left unaltered by
    the following involution:
    \begin{equation}
        \iota \colon \left[ x_{1}:x_{2}:\dots:x_{n}:x_{n+1} \right]
        \to
        \left[ x_{n}:x_{n-1}:\dots:x_{1}:x_{n+1} \right],
        \label{eq:symmcond}
    \end{equation}
    \label{def:symm}
    i.e. $\iota^{*}\left( I \right)=I$.
\end{definition}

We then have the following characterization of integrability for
autonomous recurrence relations:
\begin{enumerate}[(i)]
    \item {\bf Existence of invariants} A $n$-dimensional
        map is (super)integrable if \emph{there exists $n-1$ invariants}.
    \item {\bf Liouville integrability \cite{Veselov1991,Maeda1987,Bruschietal1991}} A $n$-dimensional
        map (in affine coordinates) is integrable if \emph{it preserves  a Poisson structure of rank $2r$
        and  $r+n-2r=n-r$ functionally independent invariants in involution with respect to this
        Poisson structure}.
        In affine coordinates 
        $\vec{w}=\left( w_{n-1},\dots,w_{0} \right)=\left[ w_{n-1}:\dots:w_{0}:1 \right]$
        we say that a map $\varphi\colon \vec{w}\mapsto\vec{w'}$
        is called a Poisson map of rank $2r\leq n$ if there is a skew-symmetric 
        matrix $J({\bf{w}})$  of rank $2r$ satisfies the Jacobian identity 
        \begin{equation}
        \label{eq:JacoIden}
        \sum_{l=1}^{n}\left(J_{li}\frac{\partial J_{jk}}{\partial w_{l-1}}
            +J_{lj}\frac{\partial J_{ki}}{\partial w_{l-1}}
            +J_{lk}\frac{\partial J_{ij}}{\partial w_{l-1}}\right)=0,
            \quad
            \forall i,j,k.
        \end{equation}
and
\begin{equation}
\label{eq:Poisson1}
\ud\varphi  J({\bf{w}})  \ud\varphi^T=J({\bf{w}}^{'}),
\end{equation}
where $\ud\varphi$ is the  Jacobian  matrix of the map $\varphi$, 
see \cite{CapelSahadevan2001, Olver1986}.
The Poisson bracket of two smooth functions $f$ and $g$ is defined as
\begin{equation}
\label{eq:Poisson_def}
\{f, g\}=\nabla f \  J \  (\nabla g)^{T},
\end{equation}
where $\nabla f$ is the gradient of $f$. We can easily see that $\{w_{i-1},w_{j-1}\}=J_{ij}$.
We note that in the case where the Poisson structure has full rank, i.e. $n=2r$, 
we only need $n/2$ invariants which are in involution.
In this case the Poisson matrix is invertible, and its inverse is called
a \emph{symplectic matrix}.
A symplectic matrix give raise to a \emph{sympectic structure}.

    \item {\bf Existence of a Lax pair \cite{Lax1968}} An $n$-dimensional
        map is integrable if it \emph{arises as compatibility condition
        of an overdetermined linear system}.
        We emphasize the fact that the Lax pair needs to provide us some integrability 
        aspects of the maps such as invariants  or solutions of the   
        non-linear system.
        It is known in the literature that not all the Lax pairs satisfy
        such conditions \cite{CalogeroNucci1991,HayButler2013,HayButler2015,GSL_Gallipoli15}.
        Lax pairs that do not satisfy such conditions are called
        \emph{fake Lax pairs} and their existence cannot be used to prove
        integrability of a given system.
    \item {\bf {Low growth condition \cite{Veselov1992,FalquiViallet1993,BellonViallet1999}}} 
        An $n$-dimensional \emph{bi-rational}
        map is integrable if \emph{the degree of growth of the iterated
        map $\varphi^{k}$ is polynomial with respect to the initial conditions
        $\pvec{x_{0}}$}.
        Integrability is then equivalent to the vanishing of the
        \emph{algebraic entropy}:
        \begin{equation}
            \varepsilon = \lim_{k\to\infty}\frac{1}{k}\log \deg_{\pvec{x_{0}}}\varphi^{k}.
            \label{eq:algentdef}
        \end{equation}
        Algebraic entropy is a measure of the \emph{complexity} of
        a map, analogous to the one introduced by Arnol'd \cite{Arnold1990}
        for diffeomorphisms.
        In this sense growth is given by computing the number of intersections of the
        successive images of a straight line with a generic hyperplane in
        complex projective space \cite{Veselov1992}. 
\end{enumerate}

We emphasize the fact  that the above list is not completely
exhaustive of all the possible definitions of integrability.
Since we are focused on autonomous recurrence relations we choose
to cover only the most used definition for these ones.

\begin{remark}
We note that algebraic entropy is invariant under bi-rational maps \cite{GrammaticosHalburdRamaniViallet2009}.
 In principle, the definition of algebraic entropy in equation
    \eqref{eq:algentdef} requires us to compute all the iterates of
    a bi-rational map $\varphi$ to obtain the sequence 
    $\left\{ d_{k}=\deg_{\pvec{x_{0}}}\varphi^{k} \right\}_{k=0}^{\infty}$.
    Fortunately, for the majority of applications the form
    of the sequence can be inferred by using generating
    functions \cite{Lando2003}:
    \begin{equation}
        g\left( z \right) = \sum_{n=0}^{\infty} d_{k}z^{k}.
    \end{equation}
    \label{eq:genfunc}
    A generating function is a predictive tool which
    can be used to test the successive members of a finite sequence.
    When a generating function is available,  the algebraic entropy is then  given by the
    logarithm of the smallest pole of the generating function,
    see \cite{GubbiottiASIDE16,GrammaticosHalburdRamaniViallet2009}.
\end{remark}

\begin{remark}
    The condition of Liouville integrability \cite{Maeda1987,Veselov1991,Bruschietal1991} is
    stronger than the existence of invariants. 
    Indeed, for a map, being measure preserving and preserving a 
    Poisson/symplectic structure are very strong conditions. 
    However, they lead to a great drop in the
    number of invariants needed for integrability.
    The same can be said for the existence of a Lax pair, since it is
    well known that a well posed Lax pair gives all the invariants of
    the system through the spectral relations.
    Finally, the low growth condition means that the complexity of
    the map is very low, and it is known that invariants help in reducing
    the complexity of a map. 
    Indeed the growth of a map possessing invariants cannot be generic
    since the motion is constrained to take place on the intersection 
    of hypersurfaces defined by the invariants.
    For maps in $\Pj^{2}$, it was proved in \cite{DillerFavre2001} that 
    the growth can be only bounded, linear, quadratic or exponential.
    Linear cases are trivially integrable in the sense of invariants.
    We note that for polynomial maps, it was already known from 
    \cite{Veselov1992} that the growth can be only linear or exponential.
    It is known that QRT mappings and other maps with invariants in $\Pj^{2}$
    possess quadratic growth \cite{Duistermaat2011book}, so the two notions
    are actually equivalent for a large class of integrable systems.
    \label{rem:intcond}
\end{remark}

Now we  discuss briefly the concept of \emph{duality} for
rational maps, which was introduced in \cite{QuispelCapelRoberts2005}.
Let us assume that our map $\varphi$ possesses $L$ independent invariants, 
i.e. $I_{j}$ for $j\in\left\{ 1,\dots,L \right\}$. 
Then we can form the linear combination:
\begin{equation}
    H = \alpha_{1} I_{1} + \dots+\alpha_{L} I_{L}.
    \label{eq:hdef}
\end{equation}
%
For an unspecified
autonomous recurrence relation
\begin{equation}
    \left[ x_{1}:x_{2}:\dots:x_{n+1} \right]
    \mapsto
    \left[ x_{1}':x_{1}:\dots:x_{n} \right]
\end{equation}
we can write down the invariant condition for $H$ \eqref{eq:hdef}:
\begin{equation}
    \widehat{H}(x_{1}',\pvec{x})=H\left(\pvec{x'}  \right) - H\left( \pvec{x} \right)
    = 0.
    \label{eq:hinvc}
\end{equation}
Since we know that $\pvec{x'}=\varphi\left( \pvec{x} \right)$ is a solution
of \eqref{eq:hinvc} we have the following factorization:
\begin{equation}
    \widehat{H}(x_{1}',\pvec{x})
    = A\left( x_{1}',\pvec{x} \right)
    B\left( x_{1}',\pvec{x} \right).
    \label{eq:hfact}
\end{equation}
We can assume without loss of generality that the map
$\varphi$ corresponds to the annihilation of $A$ in \eqref{eq:hfact}.
Now since $\deg_{x_{1}'}\widehat{H}=\deg_{x_{1}}H$ and 
$\deg_{x_{n}}\widehat{H}=\deg_{x_{n}}H$
we have that if if $\deg_{x_{1}} H,\deg_{x_{n}} H >1$ the factor
$B$ in \eqref{eq:hfact} is non constant\footnote{We remark that this
    assertion is possible because we are assuming that all the invariants
    are non-degenerate. It is easy to see that degenerate invariants can
violate this property.}.
In general, since the map $\varphi$ is bi-rational, we have the following
equalities:
\begin{subequations}
    \begin{align}
        \deg B_{x_{1}'}&=\deg_{x_{1}'}\widehat{H}-\deg_{x_{1}'}A=\deg_{x_{1}}H-1,
        \\
        \deg B_{x_{n}}&=\deg_{x_{n}}\widehat{H}-\deg_{x_{n}}A=\deg_{x_{n}}H-1.
    \end{align}
    \label{eq:degABH}
\end{subequations}
Therefore we have that if $\deg_{x_{1}}H,\deg_{x_{n}}H>2$, 
the annihilation of $B$ does not define
a bi-rational map in general, but an algebraic one.
However when  $\deg_{x_{1}} H,\deg_{x_{n}} H =2$
the annihilation of $B$ defines a bi-rational projective map.
We call this map the \emph{dual map} and we denote it by $\varphi^{\vee}$.

\begin{remark}
    We note that in principle for $\deg_{x_{1}}H=\deg_{x_{n}}H=d>2$, 
    more general factorizations can be considered:
    \begin{equation}
        \widehat{H}\left( x_{1}',\pvec{x} \right)
        = \prod_{i=1}^{d} A_{i}\left( x_{1}',\pvec{x} \right),
        \label{eq:genHfact}
    \end{equation}
    but we will not consider this case here.
    \label{rem:facthigh}
\end{remark}

Now assume that the invariants (and hence the map $\varphi$)
depends on some \emph{arbitrary constants} 
$I_{i}=I_{i}\left( \pvec{x};a_{i}\right)$, for $i=1,\dots,M$.
Choosing some of the $a_{i}$ in  such a way that
there remains $M$ arbitrary constants and such that for
a subset $a_{i_{k}}$ we can write
equation \eqref{eq:hdef} in the following way:
\begin{equation}
    H = a_{i_{1}} J_{1} + a_{i_{2}} J_{2}+\dots+a_{i_{K}}J_{a_{i_{K}}},
    \label{eq:hJdef}
\end{equation}
where $J_{i}=J_{i}\left( \pvec{x} \right)$, $i=1,2,\dots,K$
are new functions.
The parameters $a_{i_{k}}$ do not appear in the dual maps in the same way as the parameters $\alpha_i$ do not appear in the main maps. Therefore,  using the factorization \eqref{eq:hfact}
the $J_{i}$ functions are invariants for the dual maps.  

\begin{remark}
In fact, one can consider  more general combinations than  linear combinations given in \eqref{eq:hdef} and \eqref{eq:hJdef}. However, we only consider those  linear combinations  given \eqref{eq:hdef} and \eqref{eq:hJdef} in this paper.

    It is clear from equation \eqref{eq:hJdef} that even though
    the dual map is naturally equipped with some invariants, 
    it is not \emph{necessarily} equipped with a sufficient
    number of invariants to claim integrability.
    In fact there exist examples of dual maps with any possible
    behaviour, integrable, superintegrable and non-integrable
    \cite{JoshiViallet2017,GJTV_class}.
    \label{eq:dualmapints}
\end{remark}

In a recent paper \cite{JoshiViallet2017},  the authors considered the
\emph{autonomous limit} of the second member of the 
$\dP_\text{I}$ and $\dP_\text{II}$ hierarchies \cite{CresswellJoshi1999}.
We will denote these equations as $\dP_\text{I}^{(2)}$ and $\dP_\text{II}^{(2)}$ equations.
These $\dP_\text{I}^{(2)}$ and $\dP_\text{II}^{(2)}$ equations
are given by autonomous recurrence relations of order four,
and showed to be integrable
according to the algebraic entropy approach.
They showed that both maps possess two invariants, one of degree
pattern $\left( 1,3,3,1 \right)$ and one of degree pattern $\left( 2,4,4,2 \right)$. 
Using these invariants,  they showed that the dual maps of the 
$\dP_\text{I}^{(2)}$ and $\dP_\text{II}^{(2)}$ equations are
integrable according to the algebraic entropy test and moreover, produced some
invariants, showing that these dual maps were actually superintegrable.
Finally they gave a scheme to construct autonomous recurrence relations
with the assigned degree pattern $\left( 1,3,3,1 \right)$ 
associated with $\Ilow$ and $\left( 2,4,4,2 \right)$  associated with $\Ihigh$
and they provided some new examples out of this construction.

In a forthcoming paper \cite{GJTV_class} we consider the problem of finding
all fourth order bi-rational maps 
$\varphi\colon[x:y:z:u:t]\mapsto[x':y':z':u':t']$ 
possessing a polynomial a symmetric invariant 
$I_{\text{low}}$ such that $\dpat I_{\text{low}}=\left( 1,3,3,1 \right)$
\emph{where the only non-zero coefficients are those
    appearing in the $\left( 1,3,3,1\right)$ invariant of both the
    $\dP_\text{I}^{(2)}$ and  $\dP_\text{II}^{(2)}$ equation}, and 
    such that $\varphi$ possesses a polynomial 
symmetric invariant $I_{\text{high}}$  such that 
$\dpat I_{\text{high}}=\left( 2,4,4,2 \right)$.
The two invariants $\Ilow$ and $\Ihigh$ are assumed to be
functionally independent and non-degenerate.
Within this class we have found the known $\dP_\text{I}^{(2)}$ and 
$\dP_\text{II}^{(2)}$ equations as well as new examples of maps with
these properties.

In this letter we will present in detail four particular
examples of this class.
In Section \ref{sec:ex}, we will discuss two pairs of main-dual maps.
We will discuss the integrability property of these maps
in light of their invariants and of their growth.
We will present maps possessing two invariants
and integrable according to the algebraic entropy test with \emph{cubic growth}.
This implies that another rational invariant cannot exist. 
Indeed, the orbits of superintegrable maps with rational invariant
are confined to \emph{elliptic curves} and the growth is at most \emph{quadratic}
\cite{Bellon1999,Gizatullin1980}.
From this general statement follows that a four-dimensional
map with cubic growth can possess at most two rational invariants.
We note that some examples of cubic growth were already presented
in \cite{JoshiViallet2017}. 
However, it was pointed out that these examples can be deflated 
to lower dimensional maps with quadratic growth. 
This also holds for our maps, i.e. we can deflate them to integrable maps in  lower dimension.
Furthermore, we will present a map with two invariants and 
\emph{exponential growth}, that is non-integrable according to the algebraic
entropy test.
We discuss some possible reasons why this map is non-integrable even
though it possesses two invariants.
In the final Section,  we will give some conclusions
and an outlook on the future perspectives of this approach.

\clearpage

\section{Notable examples}
\label{sec:ex}

In this section  we discuss two pairs of maps,  
which arise as part of a systematic classification to be presented
in \cite{GJTV_class}.
The interest in these particular maps 
arises since the relation between their invariants and  growth properties is non trivial.
In both cases the main maps possess two functionally
independent invariants, but they behave differently. 
One map has \emph{cubic} degree growth, while the other one has
\emph{exponential} degree growth. Therefore, even though these two maps have the same number of invariants with the same degree patterns,  one map is integrable and the other one is non-integrable. 
In addition, in both cases the degree growth property of the dual maps reflect the growth
of the main map.
However, we note that the degree growth of the dual map does not always 
reflect that of  the main map\cite{GJTV_class}.


\subsection{\eqref{eq:Mi} and its dual map \eqref{eq:Di}}

Consider the map $\pvec{x}\mapsto\varphi_{i}\left( \pvec{x} \right)=\pvec{x'}$ 
given as follows:
\begin{equation}
    \begin{aligned}
        x'&
        \begin{aligned}[t]
            &=
            -\{[\nu t^2 (x+ z)+u z^2] y+t^2 \mu u z+(x+z)^2 y^2\} d-a t^4,
        \end{aligned}
        \\
        y'&=x^2 d (t^2 \mu+x y), 
        \quad
        z'=y x d (t^2 \mu+x y),
        \\
        u'&= z x d (t^2 \mu+x y), 
        \quad
        t'= t x d (t^2 \mu+x y).
    \end{aligned}
    \tag{P.i}
    \label{eq:Mi}
\end{equation}
This map depends on four parameters $a, d$ and $\mu,\nu$.

From the construction in \cite{GJTV_class} we know that the 
map \eqref{eq:Mi} possesses the following invariants:
\begin{subequations}
    \begin{align}
        t^{6}\Ilow^{\text{\ref{eq:Mi}}} &
        \begin{aligned}[t]
        &=
        a t^4 y z+d\left[ \nu y^2 z^2-y z  (u x-u z-x y) \mu\right] t^2
        \\
        &\quad
        -y^2 z^2 d (u x-x y-y z-u z),
        \end{aligned}
        \label{eq:IlowMi}
        \\
        t^{8}\Ihigh^{\text{\ref{eq:Mi}}} &
        \begin{aligned}[t]
        &=
        \left[(u z+x y-y z) \mu-\nu y  z\right] a t^6
        \\
        &
		\quad        
        +\left[y z (x y+y z+u z) a+d\mu^2 (u z+x y-y z)^2 \right.
        \\
        &
        \quad
        +\left.  2 d\mu \nu y z (u x-y z)  -d \nu^2 y^2 z^2\right] t^4
        \\
        &
        \quad
        +\left[2 d z y (u z+x y-y z) (x y+y z+u z) \mu+2 d y^2 z^2 \nu u x\right] t^2
        \\
        &
        \quad
        +d y^2 z^2 (x y+y z+u z)^2.
        \end{aligned}
        \label{eq:IhighMi}
    \end{align}
    \label{eq:intMi}
\end{subequations}

Moreover, the map  \eqref{eq:Mi} has the following degrees of iterates:
\begin{equation}
    \begin{aligned}
        \left\{d_n\right\}_{\text{\ref{eq:Mi}}}
        = &1, 4, 12, 28, 52, 86, 130, 188, 260, 348, 452,
        \\
        &
        \quad
        576, 720,886, 1074, 1288,1528, 1796, 2092\dots
    \end{aligned}
    \label{eq:degMi}
\end{equation}
The generating function of the sequence \eqref{eq:degMi} is given by:
\begin{equation}
    g_{\text{\ref{eq:Mi}}}(s) 
    =\frac{s^7-3 s^6+s^5-s^4+3 s^3+3 s^2+s+1}{(s+1)(s^2+1)(s-1)^4}.
    \label{eq:gfMi}
\end{equation}
Due to the presence of $\left( s-1 \right)^{4}$ in the denominator
we have that the growth of the map \eqref{eq:Mi} is fitted
by a \emph{cubic polynomial}.
As discussed in the Introduction this means at once that the map 
is integrable according to the algebraic entropy test and that 
another rational invariant cannot exist.  This suggests that the
geometry of the orbits of the map \eqref{eq:Mi} is nontrivial, 
and goes beyond the existence of \emph{elliptic fibrations}.

\begin{figure}[hbt]
    \centering
    \includegraphics[width=0.5\textwidth]{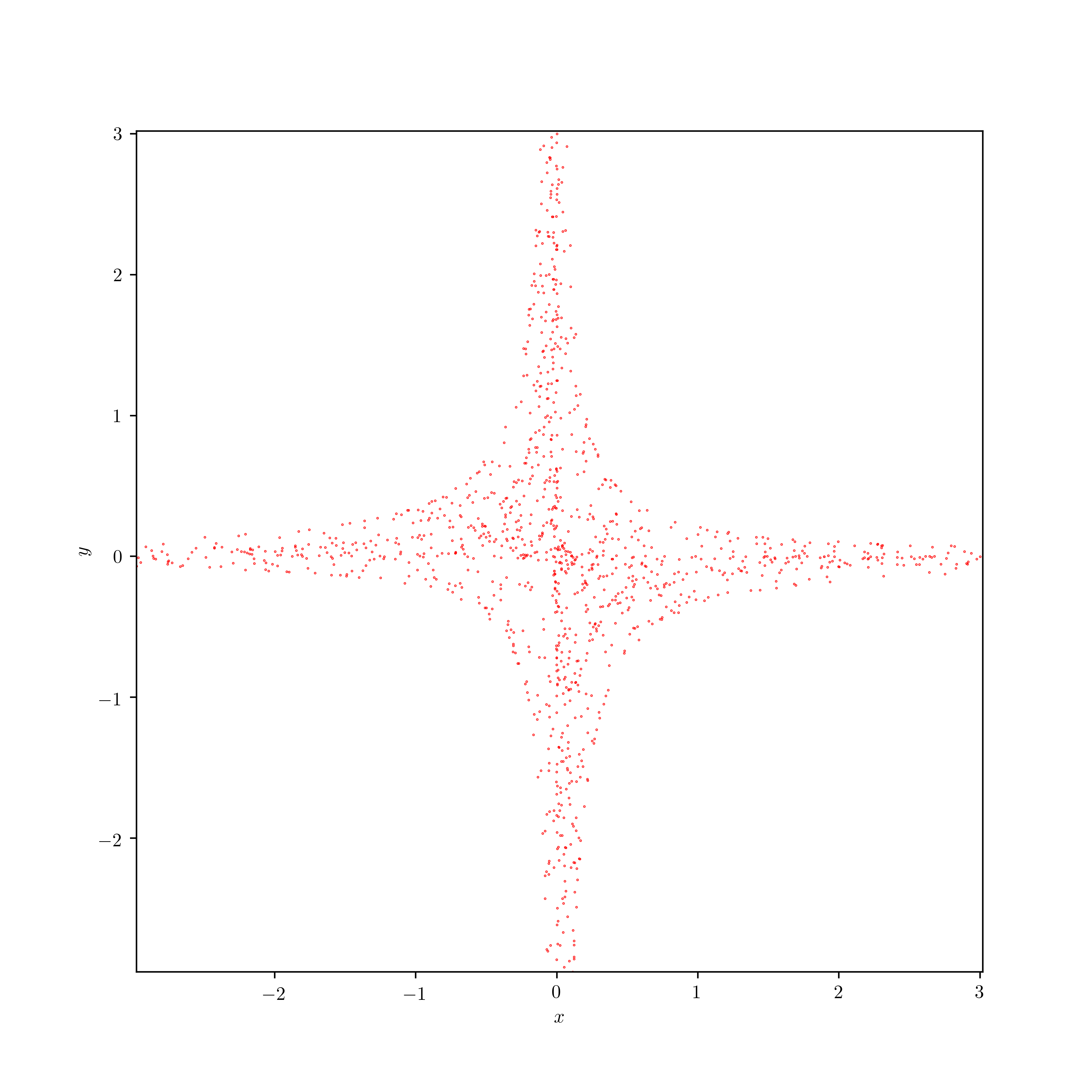}
    \caption{Affine orbit of equation \eqref{eq:Mi} with parameters
    $a=6$, $\mu=3$, $\nu=4$ and $d=6$ and initial conditions
$\left(x,y,z,u  \right)=\left( 0.02, 0.05, 0.06, 0.07 \right)$.}
    \label{fig:Pi}
\end{figure}

Explicit numerical calculations and drawings suggest that in
the case of map \eqref{eq:Mi}, no additional invariant exists.
Indeed, if an additional third invariant, even algebraic, existed
then all the orbits of of equation \eqref{eq:Mi} would lie on a curve.
On the other hand referring to Figure \ref{fig:Pi} we see that a generic
orbit of equation \eqref{eq:Mi} does not lie on a curve.
This implies that no such an invariant might exist.

The dual map $\pvec{x}\mapsto\varphi^{\vee}_{i}\left( \pvec{x} \right)=\pvec{x'}$ 
of \eqref{eq:Mi} is given by:
\begin{equation}
    \begin{aligned}
        x'&
        \begin{aligned}[t]
            &=
            \left[\beta (2 x y-2 y z+u z) \mu+(\beta \nu-\alpha) y (x-z)\right] t^2
            \\
            &
			\quad            
            +\beta y (z^2 y-x^2 y+u z^2)
        \end{aligned}
        \\
        y' &=  x^2 \beta  (t^2 \mu+x y), \quad
        z' =  y x \beta  (t^2 \mu+x y), 
        \\
        u' &= z x \beta  (t^2 \mu+x y), \quad
        t' = t x \beta  (t^2 \mu+x y).
    \end{aligned}
    \tag{Q.i}
    \label{eq:Di}
\end{equation}
This map depends on three parameters $\alpha,\beta$, and $\mu,\nu$.
The parameters $\mu$ and $\nu$ are shared with the main map \eqref{eq:Mi}.

The main map \eqref{eq:Mi} possesses two invariants and 
depends on $a$ and $d$ whereas the dual map \eqref{eq:Di} does not
depend on them. Then according to \eqref{eq:hJdef} we can write down the invariants
for the dual map \eqref{eq:Di} as:
\begin{equation}
    \alpha\Ilow^{\text{\ref{eq:Mi}}}+\beta\Ihigh^{\text{\ref{eq:Mi}}}
    =
    a\Ilow^{\text{\ref{eq:Di}}} +d \Ihigh^{\text{\ref{eq:Di}}}.
    \label{eq:derDi}
\end{equation}
Therefore, we obtain the following expressions:
\begin{subequations}
    \begin{align}
        t^{4}\Ilow^{\text{\ref{eq:Di}}} &
        \begin{aligned}[t]
            &=
            [y z \alpha+(\mu x y-y z \mu-y \nu z+\mu u z) \beta] t^2
            \\
            &
            \quad		            
            +\beta y z (x y+y z+u z),
        \end{aligned}
        \label{eq:I1i}
        \\
        t^{8}\Ihigh^{\text{\ref{eq:Di}}} &
        \begin{aligned}[t]
            &=
            \left\{\left[y^2 z^2 \nu-y z (u x-u z-x y) \mu\right] \alpha\right.
            \\
            &\quad
            +\left.\left[(u z+x y-y z)^2 \mu^2+2 y z (u x-y z) \nu \mu-\nu^2 y^2 z^2\right] \beta\right\} t^4
            \\
            &
            \quad
            +\left\{z^2 y^2 (x y+y z-u x+u z) \alpha\right.
            \\
            &\quad
            +\left.\left[2 y z (u z+x y-y z) (x y+y z+u z) \mu+2 y^2 z^2 \nu u x\right] \beta\right\} t^2
            \\
            &
            \quad
            +z^2 y^2 (x y+y z+u z)^2 \beta.
        \end{aligned}
        \label{eq:I2i}
    \end{align}
    \label{eq:intDi}
\end{subequations}
We remark that the invariant \eqref{eq:I1i}  
has degree pattern $\left( 1,2,2,1 \right)$ which differs from $\dpat I_{\text{low}}^{\rm P.i}$.

The map \eqref{eq:Di} has the following degrees of iterates:
\begin{equation}
    \begin{aligned}
        \left\{d_n\right\}_{\text{\ref{eq:Di}}}
        &= 1, 4, 12, 26, 48, 78, 118, 170, 234, 312, 406, 516,644,792\dots
    \end{aligned}
    \label{eq:degDi}
\end{equation}
with generating function:
\begin{equation}
    g_{\text{\ref{eq:Di}}}(s) =
    \frac{(s^3-2s^2-1)(s^3-s^2-s-1)}{(s^2+s+1)(s-1)^4}.
    \label{eq:gfDi}
\end{equation}
This means that the dual map is integrable according to the
algebraic entropy test with \emph{cubic} growth, just like the main map.

\begin{figure}[hbt]
    \centering
    \subfloat[][Parameters $\alpha=3$, $\mu=3$, $\nu=7$ and $\beta=3$.]{%
        \includegraphics[width=0.45\textwidth]{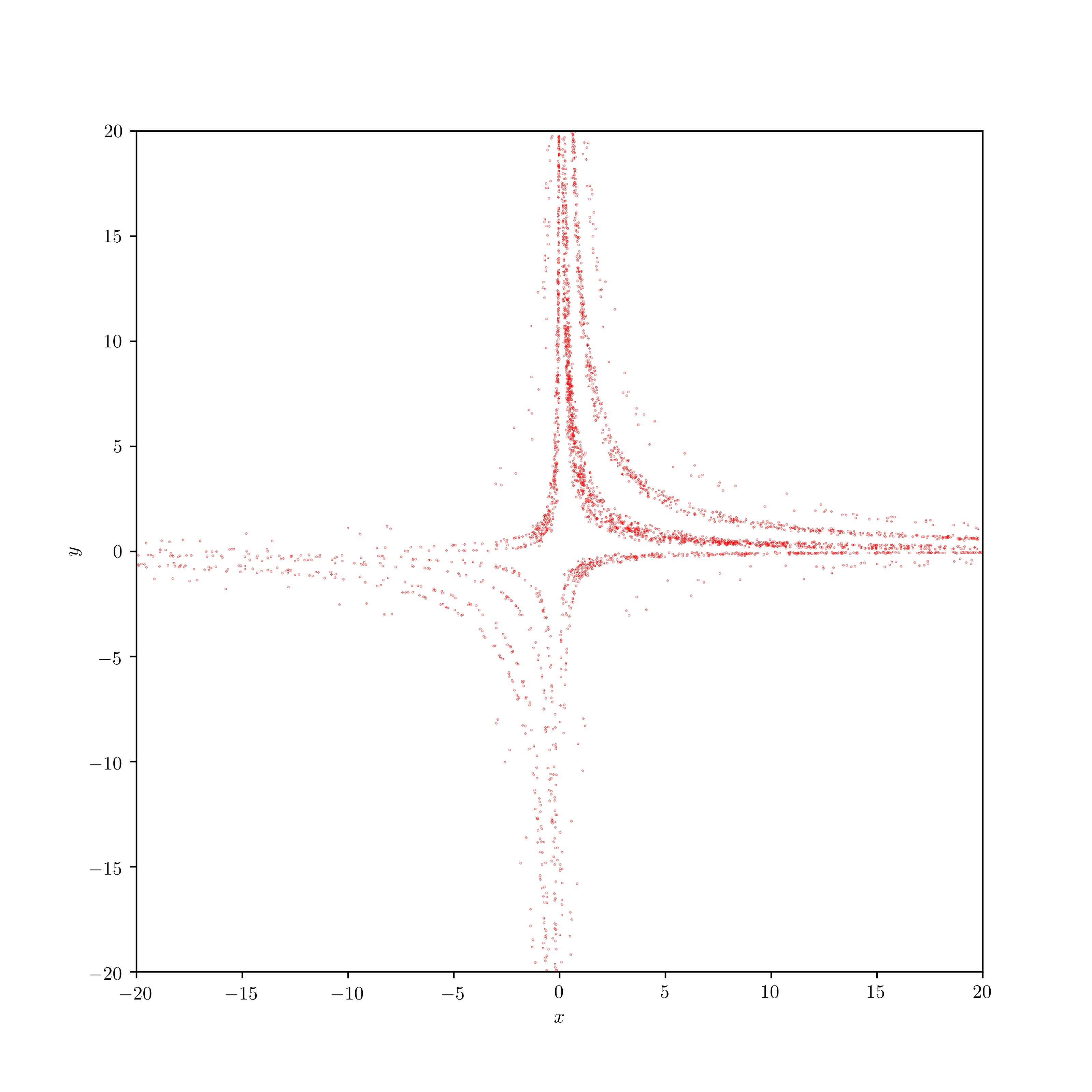}
        \label{fig:QiA}
    }
    \subfloat[][Parameters $\alpha=3$, $\mu=6$, $\nu=8$ and $\beta=9$.]{%
        \includegraphics[width=0.45\textwidth]{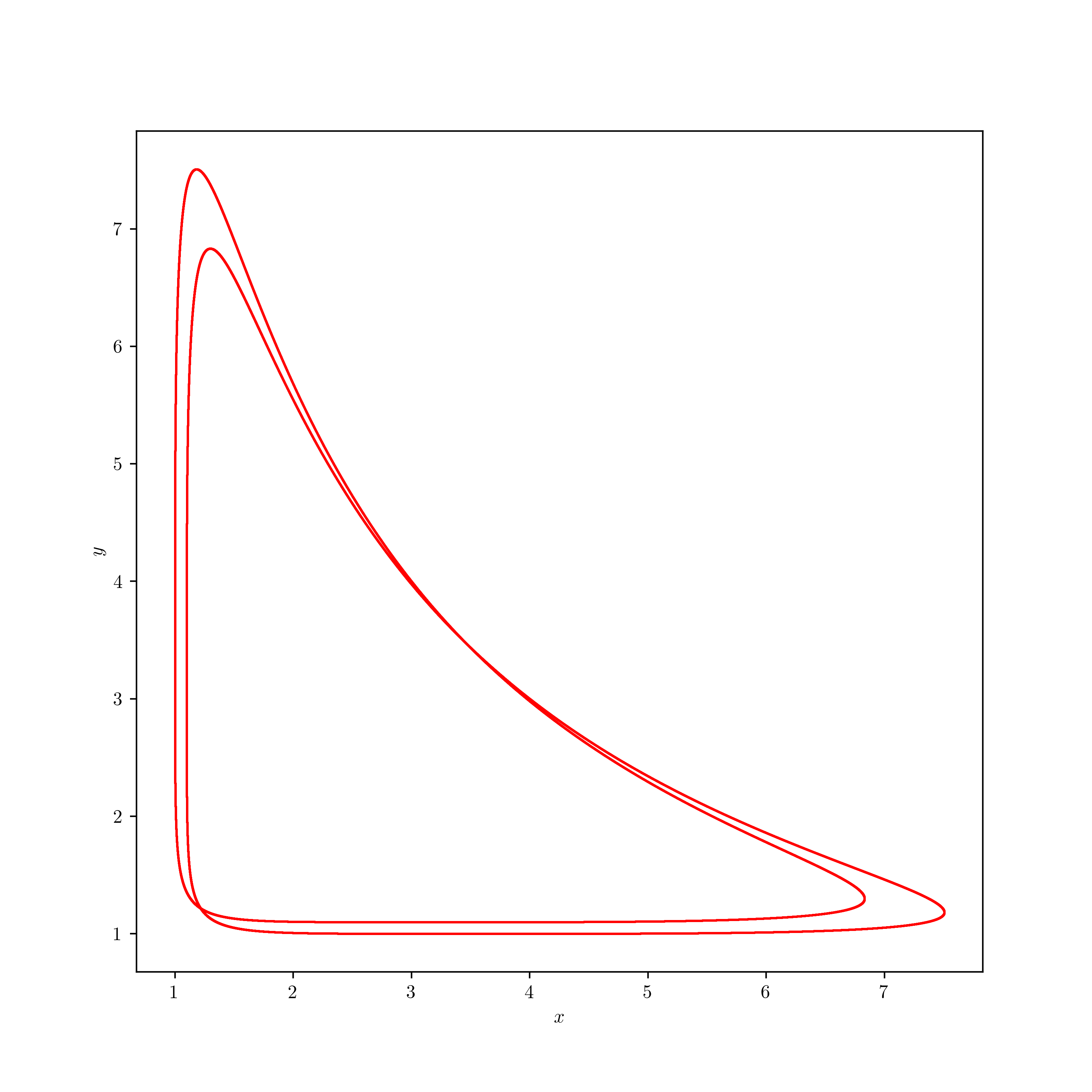}
        \label{fig:QiB}
    }
    \caption{Affine orbit of equation \eqref{eq:Di} with different
        parameters but the same initial conditions 
        $\left(x,y,z,u  \right)=\left( 3, 4, 1, 3 \right)$.}
    \label{fig:Qi}
\end{figure}

Explicit numerical calculations and drawings suggest that also in
the case of map \eqref{eq:Di}, no additional invariant exists.
Indeed, if an additional third invariant, even algebraic, existed
then all the orbits of of equation \eqref{eq:Di} would lie on a curve.
In this case we are actually able to find some orbits lying on a curve,
see Figure \ref{fig:QiB}. 
However, it is possible to find orbits of equation \eqref{eq:Di} that 
do not lie on a curve.
An example of such orbit is shown in Figure \ref{fig:QiA}. 
Therefore, we can conclude that a \emph{globally defined} third invariant 
does not exist.
The existence of some closed orbits like in Figure \ref{fig:QiB}
suggest the existence of a non-analytic invariant existing only in some
regions of the space.

Therefore,  the pair of main-dual maps \eqref{eq:Mi} and \eqref{eq:Di}
consists of two integrable equations with non-standard degree of growth.
However,  as remarked above the degree pattern of the invariants
of the maps \eqref{eq:Mi} and \eqref{eq:Di} differ slightly.

We now consider the  maps   \eqref{eq:Mi} and \eqref{eq:Di} in 
affine coordinates  which are given by
\begin{equation}
\label{eq:Affine_map}
\varphi:(w_3,w_2,w_1,w_0)\mapsto (w_4, w_3,w_2,w_1),
\end{equation}
where 
\begin{align}
    \tag{AP.i}
    w_4&=\frac{N_1}{d w_{{3}}\left( w_{{2}}w_{{3}}+\mu \right) },
    \\
    \tag{AD.i}
    w_4&=\frac{N_2}{\beta \ w_{{3}}  \left( w_{{2}}w_{{3}}+\mu \right) },
\end{align}
with
\begin{align}
    N_1&
    \begin{aligned}[t]
        &= - d \big( w_{0}w_{1}^{2}w_{2}+w_{1}^{2}w_{2}^{2}+2
        w_{1}w_{2}^{2}w_{3}+w_{2}^{2}w_{3}^{2}+\mu w_{0}w_{1}
        \\
        &\quad+\nu w_{1}w_{2}+\nu w_{2}w_{3} \big) - a,
    \end{aligned}
    \\
    N_2&
    \begin{aligned}[t]
        &=\beta w_{0}w_{1}^{2}w_{2}+\beta \mu w_{0}w_{1}+\beta 
        w_{1}^{2}w_{2}^{2}+ \left(\alpha -2 \beta \mu-\beta \nu\right) w_{1}w_{2}
    \\
     &\quad 
     -\beta w_{3}^{2}w_{2}^{2}+ \left( 2 \beta \mu +\beta \nu-\alpha \right) w_{2}w_{3}.
    \end{aligned}
\end{align}
Invariants for these  maps are obtained  from $\Ilow$ and $\Ihigh$ 
respectively by taking $t=1, u=w_0, z=w_1,y=w_2,$ and $x=w_3$.

We note that when a Poisson structure has the full rank, 
using equation \eqref{eq:Poisson1}, one gets  
\begin{equation}
    \big[\det(\ud\varphi)\big]^2=\frac{\det\big(J({\bf{w}}^{'})\big)}{\det\big(J({\bf{w}})\big)}.
\end{equation}
This implies that the map $\varphi$ is either volume or anti-volume preserving.

We recall that a map $\varphi$ is called (anti)volume preserving 
if there is a function $\Omega(\vec{w})$ such that
the following volume form is preserved
\begin{equation}
\label{eq:volume_form}
    \Omega(\vec{w}) \ud w_0\wedge \ud w_1\wedge\ldots \wedge \ud w_{n-1}
    = \pm \Omega(\vec{w'}) \ud w_0'\wedge \ud w_1'\wedge \ldots \wedge \ud w_{n-1}'.
\end{equation}
Thus, we can  write 
\begin{equation}
\label{eq:volume_preserve}
\frac{\partial\big(w_0', w_{1}',\ldots, w_{n-1}'\big)}{\partial\big(w_0, w_{1},\ldots, w_{n-1}\big)}
=\pm \frac{\Omega({\bf{w}})}{\Omega({\bf{w}}')},
\end{equation}
where the left hand side is the determinant of the Jacobian matrix of the map $\varphi$.
In \cite{ByrnesHaggarQuispel1999} it was proved that
if a map  in $n$ dimension is (anti) volume preserving 
and possesses $n-2$ invariant, then we can construct an (anti) Poisson 
structure of rank 2  from these invariants.  
However, these invariants turn out to be Casimirs 
(functions that Poisson commute with all other functions) 
of this Poisson bracket. 
Therefore, in order to have Liouville integrability we need 
an extra invariant apart from the known $n-2$ invariants 
if we want to use use Poisson structures constructed this way. 
In  other words,  the map  is super integrable.  
Thus, to  discuss about Liouville  integrability of 
the maps (AP.i) and (AQ.i) we need to find  a Poisson bracket of rank $4$ as we already predicted that the third invariant does not exist.
We do not have that information for these maps but we can show 
they reduce to three dimensional Liouville integrable 
maps via a process called deflation \cite{JoshiViallet2017}.
\emph{Mutatis mutandis}, this process will preserve the invariants,
and in dimension three two invariants are sufficient to claim integrability in the general
sense as discussed in the Introduction.

It is easy to check that the maps (AP.i) and (AQ.i) are volume and anti-volume
preserving, respectively, with respect to the same volume form:
\begin{equation}
\label{eq:OmegaPQi}
\Omega=w_1w_2(w_1w_2+\mu).
\end{equation}
We now construct the (anti) Poisson structures for these two maps 
following \cite{ByrnesHaggarQuispel1999}. We consider the  dual multi-vector of the volume form
\begin{equation}
\label{eq:dual_volume_form}
\tau=m \frac{\partial}{w_0}\wedge \frac{\partial}{w_1}\wedge \frac{\partial}{w_2}\wedge \frac{\partial}{w_3},
\end{equation}
where $m=1/\Omega$. 
A degenerate Poisson structure  for the map (AP.i)  and a degenerate  
anti-Poisson structure for the map (AQ.i) are  given by the following contraction
\begin{equation}
\label{eq:construct_Poisson}
J=\tau \rfloor \ud\Ilow \rfloor \ud\Ihigh, 
\end{equation}
where $\Ilow$ and $\Ihigh$ are invariants for these maps in affine coordinates. 
Since these (anti) Poisson structures are quite big,  
we do not present them here. 

\begin{remark}
    The Poisson structures which can be constructed using the
    method of \cite{ByrnesHaggarQuispel1999} are degenerate and
    cannot be used to explain the integrability of the two maps
    (AP.i) and (AQ.i).
    \label{rem:degpoisson}
\end{remark}

We also note that the maps (AP.i) and (AQ.i) can be reduced to 
three dimensional maps  using a deflation $v_i=w_iw_{i+1}$. 
The recurrences for these maps are denoted by (DP.i) and (DQ.i) 
and are given as follows
\begin{align*}
\tag{DP.i}&d\mu \left(v_{0}+v_{3}\right)+d\nu \left( v_{1}+v_{2}\right)+d\left(v_{0}v_{1}
+v_{1}^{2}+2v_{1}v_{2}+v_{2}^{2}+v_{2}v_{3}\right)+a
=0,\\
\tag{DQ.i}&
\beta\mu \left(-v_{0}+2\,\beta v_{1}-2\,\beta v_{2}+
v_{3}\right)+ (\beta\nu-\alpha) \left(v_{1}-v_{2}\right)
\\
&\quad
+\beta\left(-v_{0}v_{1}
 -v_{1}^{2}+v_{2}^{2}+v_{2}v_
{3}\right)=0.
\end{align*}
Each of the maps  (DP.i) and (DQ.i) has two functionally 
independent invariants which can be  obtained directly from $\Ilow$ and $\Ihigh$
even though they live in a different space.
One can check that the map (DP.i) and (DQ.i) are anti-volume  preserving 
and volume preserving with $\Omega=v_1+\mu$. 
Therefore,  we can construct their (anti) Poisson structure using 
the three dimensional version of  \eqref{eq:construct_Poisson}.
Using the following  invariant from $\Ilow$ for (DP.i)
\begin{equation}
\label{eq:dPi_INV}
I_1^{\rm P.i}=d\mu v_{{0}}v_{{1}}-d\mu v_{{0}}v_{{2}}+d\mu v_{{1}}v_{{2}}+d\nu {
v_{{1}}}^{2}+dv_{{0}}{v_{{1}}}^{2}-dv_{{0}}v_{{1}}v_{{2}}+d{v_{{1}}}^{
3}+d{v_{{1}}}^{2}v_{{2}}+av_{{1}}
\end{equation}
 we have found that the map
(dP.i) has an anti-Poisson structure given by
\begin{align*}
J_{12}^{\rm P.i}&=d(v_1-v_0), \ J_{2,3}^{\rm P.i}=d(v_1-v_2)\\
J_{13}^{\rm P.i}&={\frac {-d\mu v_{{0}}-d\mu v_{{2}}-2 d\nu v_{{1}}-2 dv_{{0}}v_{{1
}}+dv_{{0}}v_{{2}}-3 d{v_{{1}}}^{2}-2 dv_{{1}}v_{{2}}-a}{\mu+v_{{1}}
}}.
\end{align*}

Similarly, for the map (DQ.i) we obtain the invariant
\begin{equation}
\label{eq:dQi_INV}
I_1^\text{DQ.i}=\beta \mu v_{{0}}-\beta \mu v_{{1}}+\beta \mu v_{{2}}-\nu \beta
 v_{{1}}+\beta v_{{0}}v_{{1}}+\beta {v_{{1}}}^{2}+\beta v_{{1}}v_{
{2}}+\alpha v_{{1}},
\end{equation}
and the corresponding Poisson structure
\begin{equation}
\label{eq:dQi_Poisson}
J^{\rm Q.i}=
    \left[ 
        \begin {array}{ccc} 
            0 & \beta & {\dfrac {\beta \left(\mu+\nu-v_{{0}}-2 v_{{1}}-v_{{2}}\right)-\alpha}{\mu+v_{{1}}}}
            \\ 
            -\beta & 0 & \beta
            \\ 
            -{\dfrac {\beta \left(\mu+\nu-v_{{0}}-2 v_{{1}}-v_{{2}}\right)-\alpha}{\mu+v_{{1}}}}&-\beta&0
        \end {array} 
    \right].
\end{equation}
For these constructions,  $ I_1^{\rm P.i}$ and  $ I_1^{\rm Q.i}$ 
are  Casimirs for  their associated (anti) Poisson structures. 
Their second (anti) Poisson structures can be  obtained from 
the invariant $\Ihigh$ but we do not present here as they are quite big.

It is important to note that  the (anti) Poisson structures of (AP.i)and (AQ.i) 
under inflation  give us the  trivial Poisson structures for (DP.i) and (DQ.i), 
i.e. $J={\bf 0}$, where ${\bf 0}$ is the zero matrix. 
On the other hand, from the common factor that appears in the Poisson structure 
of (AP.i), we have found  that there exists  an anti-invariant $K^{\rm P.i}$ 
for this map, i.e.  $K^{\rm P.i}({\bf w})=-K^{\rm P.i}({\bf w'})$where 
\begin{equation}
\label{eq:anti-integral1}
K^{\rm P.i}=2d\ \big(w_{2}w_{1}^{2}w_{0}+w_{2}^{2}w_{1}^{2}+w_{1}w_{2}
^{2}w_{3}+\mu w_{0}w_{1}-\mu w_{1}w_{2}+\mu w_{2}w_{3
}+\nu w_{1}w_{2}
\big)+a. 
\end{equation}
However, $K^{\rm P.i}$ is not  independent of  
$I^{\rm P.i}_{low}$ and $I^{\rm P.i}_{high}$ since we have 
\begin{equation}
\label{eq:relation}
\left(K^{\rm P.i}\right)^2-4d \ \Ihigh^{\rm P.i}- 8d\ \nu  \Ilow^{\rm P.i}=a^2.
\end{equation}
Using this anti-invariant, we obtain the following  anti-invariant for the map (DP.i)
\begin{equation}
\label{eq:dPi_antiINV}
K^{\text{DP.i}}=2 d\mu v_{{0}}-2 d\mu v_{{1}}+2 d\mu v_{{2}}+2 d\nu v_{{1}}+2
 dv_{{0}}v_{{1}}+2 d{v_{{1}}}^{2}+2 dv_{{1}}v_{{2}}+a.
\end{equation}
Therefore, using this anti-invariant, we get a Poisson structure 
for (DP.i) as follows (after factoring out a constant term) 
\begin{equation}
\label{eq:dPi_Poisson2}
J^{\rm P.i}_2=\left[ \begin {array}{ccc} 0&1&{\dfrac {\mu-\nu-v_{{0}}-2 v_{{1}}-v_{
{2}}}{\mu+v_{{1}}}}
    \\ 
    -1&0&1
    \\ 
    -{\dfrac {\mu-\nu-v_{{0}}-2 v_{{1}}-v_{{2}}}{\mu+v_{{1}}}}&-1&0
\end {array} \right]. 
\end{equation}
We can check directly that the invariants  inherited from the affine map (AP.i)
are in involution with respect to the Poisson structure~\eqref{eq:dPi_Poisson2}.
In the sense of the definition given in the Introduction, this
means that the reduced maps (DP.i) and (DQ.i) are Liouville integrable.

\begin{remark}
We notice that we can always use the invariants \eqref{eq:dPi_INV} 
and \eqref{eq:dQi_INV} to reduce the  three dimensional maps (DP.i) 
and (DQ.i) to two dimensional maps and relate them to QRT maps.
To be more specific we have that the reduced map of (DQ.i) preserves 
a bi-quadratic curve so that it is  of  the QRT type. 
On the other hand, using the anti-invariant, the reduced map of (DP.i) 
sends a bi-quadratic to another bi-quadratic  and fits in the framework of 
\cite{RJ15}.
\end{remark}

\subsection{\eqref{eq:Pii} and its dual map \eqref{eq:Qii}}

Consider the map $\pvec{x}\mapsto\varphi_\text{ii}\left( \pvec{x} \right)=\pvec{x'}$ 
given as follows:
\begin{equation}
    \begin{aligned}
        x' &=\left[(x^2+z^2) y-u z^2\right] \mu-t^2 (u-2y), 
        \\
        y' &= x (t^2+\mu x^2), 
        \quad
        z' = y (t^2+\mu x^2),
        \\
        u' &= z (t^2+\mu x^2), 
        \quad
        t'= t (t^2+\mu x^2).
    \end{aligned}
    \tag{P.ii}
    \label{eq:Pii}
\end{equation}
This map only depends on the parameter $\mu$.

From the construction in \cite{GJTV_class} we know
that the map \eqref{eq:Pii} has the following invariants:
\begin{subequations}
    \begin{align}
        t^{5}\Ilow^{\text{\ref{eq:Pii}}} &
        \begin{aligned}[t]
        &=\left( x-z \right)\left( u-y \right)
        \left( {t}^{2}+{z}^{2}\mu \right)  \left( \mu{y}^{2}+{t}^{2} \right),    
        \end{aligned}
        \label{eq:IlowPii}
        \\
        t^{6}\Ihigh^{\text{\ref{eq:Pii}}} &
        \begin{aligned}[t]
        &=
        \left[  \left( x-z \right) ^{2}{y}^{4}+{y}^{2}{z}^{4}-2 y{z}^{4}u+{u}^{2}{z}^{4} \right] {\mu}^{2}
        \\
        &
        \quad
        +2 {t}^{2} \left[  \left( {x}^{2}-2 xz+2 {z}^{2} \right) {y}^{2}-2 y{z}^{2}u+{u}^{2}{z}^{2} \right] \mu
        \\
        &
        \quad
        +{t}^{4} \left( {z}^{2}+{u}^{2}+{x}^{2}+{y}^{2}-2 uy-2 xz \right).
        \end{aligned}
        \label{eq:IhighPii}
    \end{align}
    \label{eq:intPii}
\end{subequations}

Moreover, the map \eqref{eq:Pii} has the following degrees of iterates:
\begin{equation}
    \begin{aligned}
        \left\{d_n\right\}_{\text{\ref{eq:Pii}}}
        &=1,3,9,21,45,93,189,381,765,1533\dots
    \end{aligned}
    \label{eq:degPii}
\end{equation}
with generating function:
\begin{equation}
    g_{\text{\ref{eq:Pii}}}(s) = \frac{1+2 s^2}{(2 s-1)(s-1)}.
    \label{eq:gfPii}
\end{equation}
This means that despite the existence of the two invariants
\eqref{eq:intPii} the map \eqref{eq:Pii} is non-integrable 
according to the algebraic entropy test: its entropy is positive 
and given by $\varepsilon=\log2$.

Therefore we have that the map \eqref{eq:Pii} is an example
of non-integrable admitting two invariants.

Again following \cite{ByrnesHaggarQuispel1999} we can produce a Poisson 
structure of rank $2$ for \eqref{eq:Pii} as the affine version of (P.ii) is volume preserving with
$\Omega=(1+\mu w_1^2)(1+\mu w_2^2)$,
where we have taken $t=1, u=w_0, z=w_1,y=w_2,$ and $x=w_3$.   
By the construction, the two invariants
\eqref{eq:intPii} become  Casimir functions for it, so again the
existence of such Poisson structure does not imply any form of
Liouville integrability.
However, we notice that there are common factors appear 
at every non-zero entries of this structure. 
Thus,  we have found the following anti-invariant for 
the map (P.ii) using these common factors
\begin{equation}
    \begin{aligned}
        K^{\rm P.ii}&=\left[
            \mu \left(w_{{0}}w_{{1}}^{2}- {w_{{1}}}^{2}w_{{2}}- w_{{1}}w_{{2}}^{2}
            + w_{{2}}^{2}w_{{3}}\right)+w_{{0}}-w_{{1}}-w_{{2}}+w_{{3}}    
        \right]\times 
        \\
        &\phantom{=\times}
        \left[
            \mu \left(w_{{0}}w_{{1}}^{2}- w_{{1}}^{2}w_{{2}}+w_{{1}}w_{{2}}^{2}
            -w_{{2}}^{2}w_{{3}}\right)+w_{{0}}+w_{{1}}-w_{{2}}-w_{{3}}
        \right]
        \\
        &=F_1F_2   
    \end{aligned}
    \label{eq:Pii_antiINV1}
\end{equation}
This suggests that we should check each factor of $K^{P.ii}$ 
to see whether they are (anti) invariants of (Pii). 
By direct calculation we can see that the first factor $F_1$ 
is an anti-invariant and $F_2$ is an invariant for (P.ii), 
but they are not functionally independent of $\Ilow$ and $\Ihigh$. 
In fact, their relations are 
    \begin{equation}
    \label{eq:Pii-INV_relation}
\Ihigh^{\rm P.ii}-F_1^2+2\Ilow^{\rm P.ii}=0, \ \mbox{and}\  \Ihigh^{\rm P.ii}-F_2^2-2\Ilow^{\rm P.ii}=0.   
    \end{equation}
Therefore, the map (P.ii) actually has two invariants of degrees 
$(1,2,2,1)$ and $(1,3,3,1)$.
Nevertheless, despite the existence of such invariants the
map (P.ii) is non-integrable in the sense of the algebraic entropy.

\begin{remark}
    We can use $F_1$ and $F_2$ to construct an anti-Poisson structure for 
    (P.ii) using the formula~\eqref{eq:construct_Poisson}
    \begin{equation}
        \begin{gathered}
    J_{1,2}=-1,\ J_{2,3}=1,\ J_{3,4}=-1\\
    J_{1,3}={\frac {2\mu w_{{1}} \left( w_{{2}}-w_{{0}} \right) }{\mu w_{{1}}^{2}+1}}, 
    \
    J_{2,4}=-{\frac { 2\mu w_{{2}}\left( w_{{3}}-w_{{1}} \right) }{\mu w_{{2}}^{2}+1}}
    \\
    J_{1,4}=-{\frac {{\mu}^{2}w_{1}w_{2} \left[ 4 \left(w_{{0}}w_{{1}}
                -w_{{0}}w_{{3}}+ w_{{2}}w_{{3}}\right)
                -3 w_{{1}}w_{{2}} \right]+\mu\left( w_{{1}}^{2}+ w_{{2}}^{2}\right)+1}{
    \left( \mu w_{{1}}^{2}+1 \right)  \left( \mu w_{{2}}^{2}+1\right) }}
        \end{gathered}
 \end{equation}
 We have checked that $F_2$ and $\Ilow^{P.ii}$ are in involution with respect to this anti-Poisson structure. 
 A Poisson structure can be obtained by multiplying this anti-Poisson structure with the anti-invariant $F_1$.
\end{remark}

The dual map $\pvec{x}\mapsto\varphi^{\vee}_{ii}\left( \pvec{x} \right)=\pvec{x'}$ 
of \eqref{eq:Pii} is given as follows:
\begin{equation}
    \begin{aligned}
        x'&
        \begin{aligned}[t]
        &= 
        \alpha  \left[  \left( {x}^{2}-{z}^{2} \right) y+u{z}^{2} \right] {\mu}
        +  {t}^{2}\alpha u+\beta {y}^{2} \left( x-z \right)  \mu
        \\&\quad
        +{t}^{2}\beta  \left( x-z \right), 
        \end{aligned}
        \\
        y' &= \alpha x  \left( {t}^{2}+\mu{x}^{2} \right),
        \quad
        z' = \alpha y  \left( {t}^{2}+\mu{x}^{2} \right), 
        \\
        u' &= \alpha z  \left( {t}^{2}+\mu{x}^{2} \right),
        \quad
        t' = \alpha t  \left( {t}^{2}+\mu{x}^{2} \right).
    \end{aligned}
    \tag{Q.ii}
    \label{eq:Qii}
\end{equation}
This map depends on three parameters $\alpha,\beta$ and $B$.
The parameter $\mu$ is shared with the main map \eqref{eq:Pii}.

Since the main map \eqref{eq:Pii} possesses two invariantss depending only  
on one parameter $\mu$ then according to \eqref{eq:hJdef} 
we can write down only a single invariant for the dual map \eqref{eq:Qii}:
\begin{equation}
    I^{\text{\ref{eq:Qii}}}=
    \alpha\Ihigh^{\text{\ref{eq:Pii}}}+\beta\Ilow^{\text{\ref{eq:Pii}}}.
    \label{eq:derQii}
\end{equation}
The invariant \eqref{eq:derQii} has degree pattern
$\left( 2,4,4,2 \right)$.

We have then that the dual map \eqref{eq:Qii} 
has the following fast-growing  degrees of iterates:
\begin{equation}
    \begin{aligned}
        \left\{d_n\right\}_{\text{\ref{eq:Qii}}}
        &=1, 3, 9, 21, 45, 93, 189, 381, 765, 1533, 3069\dots.
    \end{aligned}
    \label{eq:degQii}
\end{equation}
The growth \eqref{eq:degQii} is clearly exponential
and its generating function is
\begin{equation}
    g_{\text{\ref{eq:Qii}}}(s) = \frac{1+2 s^2}{(2 s-1)(s-1)}.
    \label{eq:gfQii}
\end{equation}
This confirms that the  algebraic entropy 
is positive and equal to $\varepsilon=\log2$.
%

This means that the dual map is non-integrable with same
rate of growth as the main map.
In this case we can show that the map is anti-volume preserving
with the same measure as the main map (P.ii). 
Moreover, we proved that the map \eqref{eq:Qii} do not possesses 
any addition invariant up to order 14.
Therefore at the present stage we cannot construct 
a Poisson structure using  the method of 
\cite{ByrnesHaggarQuispel1999}.

\section{Conclusions and outlook}
\label{sec:concl}

In this letter,  we gave some examples of fourth order bi-rational maps
with two invariants possessing interesting degree growth properties. 
These examples come from our forthcoming  
classification of all the fourth-order autonomous recurrence relations possessing two
invariants in a  given class of degree patterns \cite{GJTV_class} .

The first pair of bi-rational maps is
given by the map \eqref{eq:Mi} and its dual \eqref{eq:Di}
and consists of integrable maps with cubic growth.
The interest in maps with cubic growth arises from geometrical considerations:
maps with polynomial but higher than quadratic growth, can arise only in
dimension greater than two \cite{DillerFavre2001} and prove,
in the case of superintegrable maps, the existence
of non-elliptic fibrations of invariant varieties \cite{BellonViallet1999}.
The interest in maps with this type of growth arose recently following
the examples given in \cite{JoshiViallet2017} and we expect  them  
to lead to  many new and interesting geometric structures.

The second pair of fourth order bi-rational maps  
given by the map \eqref{eq:Pii} and its dual \eqref{eq:Qii}, 
consists of non-integrable maps with exponential growth.
There are various possible reasons why the map \eqref{eq:Pii} is
non-integrable despite possessing two invariants. 
To claim integrability with two invariants according to the
discrete Liouville theorem \cite{Maeda1987,Bruschietal1991,Veselov1991} we need to prove
that the map has a symplectic structure and that the two invariants 
commute with respect to this symplectic structure.
Hence, either the map \eqref{eq:Pii} does not admit \emph{any} symplectic
structure, or the map \eqref{eq:Pii} admits only symplectic
structures such that the two invariants \eqref{eq:intPii} do not commute.
Since, usually, from a set of non-commuting invariants it is possible to find
a set of functionally independent commuting invariants we are more leaned
to conjecture that equation \eqref{eq:Pii} is devoid of a non-degenerate
Poisson structure.

Work is in progress to characterize the surfaces generated
by the invariants in both integrable and non-integrable cases.
We expect this to give new results in the geometric theory of integrable
systems.

\bibliographystyle{plain}
\bibliography{bibliography}

\end{document}